\documentclass{ws-procs9x6}

\begin{document}

\title{Conformality and Unification of Gauge Couplings}

\author{P.H. FRAMPTON}

\address{Department of Physics and Astronomy,\\
University of North Carolina,\\
Chapel Hill, NC 27599-3255, USA.\\
E-mail: frampton@physics.unc.edu}


\maketitle

\abstracts{
By use of the AdS/CFT correspondence on orbifolds, models
are derived which can contain the standard model
of particle phenomenology. It will be assumed that the theory becomes
conformally invariant at a renormalization-group
fixed-point in the TeV region. A recent application to
TeV unification is briefly mentioned.}

\maketitle

\section{Introduction}

\bigskip

Conformality is inspired by superstring duality and assumes
that the particle spectrum of the standard model is enriched such that there
is a conformal fixed point of the renormalization group at the TeV scale.
Above this scale the coupling do not run so the hierarchy is nullified.

Until very recently, the possibility of testing string theory seemed at best
remote. The advent of $AdS/CFT$s and large-scale string compactification
suggest this point of view may be too pessimistic, since both could lead to $%
\sim 100TeV$ evidence for strings. With this thought in mind, we are
encouraged to build $AdS/CFT$ models with realistic fermionic structure, and
reduce to the standard model below $\sim 1TeV$.

Using AdS/CFT duality, one arrives at a class of gauge field theories of
special recent interest. The simplest compactification of a ten-dimensional
superstring on a product of an AdS space with a five-dimensional spherical
manifold leads to an N=4 $SU(N)$ supersymmetric gauge theory, well
known to be conformally invariant\cite{mandelstam}. By replacing the
manifold $S^5$ by an orbifold $S^5/\Gamma$ one arrives at less
supersymmetries corresponding to  $N = 2,~1 ~{\rm or}~ 0$ depending
\cite{KS} on whether $\Gamma \subset SU(2), ~~ SU(3), ~~{\rm or}
\not{\subset }SU(3)$ respectively, where $\Gamma$ is in all cases a subgroup of
$SU(4) \sim SO(6)$ the isometry of the $S^5$ manifold.

It was conjectured in \cite{maldacena} that such $SU(N)$ gauge theories are
conformal in the $N \rightarrow \infty$ limit. In \cite{F1} it was
conjectured that at least a subset of the resultant nonsupersymmetric $ N = 0$ theories are conformal even for finite $N$. Some first steps to
check this idea were made in \cite{WS}. Model-building based on abelian $%
\Gamma$ was studied further in \cite{CV,F2,F3}, arriving in \cite{F3} at an $%
SU(3)^7$ model based on $\Gamma = Z_7$ which has three families of chiral
fermions, a correct value for ${\rm sin}^2 \theta$ and a conformal scale $%
\sim 10$~~TeV.

The case of non-abelian orbifolds bases on non-abelian $\Gamma$ has not
previously been studied, partially due to the fact that it is apparently
somewhat more mathematically sophisticated. However, we shall show here that
it can be handled equally as systematically as the abelian case and leads to
richer structures and interesting results.

In such constructions, the cancellation of chiral anomalies in the
four-dimensional theory, as is necessary in extension
of the standard model ({\it e.g.} \cite{chiral,331}),
follows from the fact that the progenitor ten-dimensional
superstring theory has cancelling hexagon anomaly\cite{hexagon}.
It offers a novel approach to family unification\cite{guts,Pak}.

\bigskip

\section{Gauge Coupling Unification}

\bigskip

There is not space here to describe many technical details which are,
however, available
in the published papers cited at the end of this talk. But I would like to emphasize one
success of the approach which involves the unification of gauge couplings
\cite{F2,FMS}. Recall that the successful such unification is one primary
reason for belief in supersymmetric
grand unification {\it e.g.} \cite{Wil}. That argument is simple to state:
The RG equations are:

\begin{equation}
\frac{1}{\alpha_i(M_G)} = \frac{1}{\alpha_i(M_Z)} - \frac{b_i}{2 \pi} \ln
\left( \frac{M_G}{M_Z} \right)
\end{equation}
Using the LEP values at the Z-pole as $\alpha_3 = 0.118 \pm 0.003$,
$\alpha_2 = 0.0338$ and $\alpha_1 = \frac{5}{3}\alpha_Y = 0.0169$
(where the eroors on $\alpha_{1,2}$ are less than 1\%) and
the MSSM values $b_i = (6\frac{3}{5}, 1, -3)$ leades
to $M_G = 2.4 \times 10^{16}$ GeV and the prediction that
$\sin^2\theta = 0.231$
in excellent agreement with experiment.

In the present approach the three gauge couplings $\alpha_{1,2,3}$
run up to $\sim 1$TeV where they freeze and embed in a larger (semi-simple)
gauge group which contains $SU(3) \times SU(2) \times U(1)$.

I will give two examples, the first based on the abelian orbifold $S^5/Z_7$
and the second based on the non-abelian orbifold $S^5/(D_4 \times Z_3)$.

In the first, abelian, case we choose N=3, $\Gamma=Z_7$
and the unifying group\cite{F2,F3} is therefore $SU(3)^7$.
It is natural to accommodate one $SU(3)$ factor (color)
into one of the seven $SU(3)$ factors, $SU(2)_L$ as a diagonal subgroup
of two
and to identify the correctly normalized $U(1)$
as the diagonal subgroup of the remaining four
$SU(3)$ factors. This implies that $\alpha_2/\alpha_1 = 2$
and consequently:

\begin{equation}
\sin^2\theta = \frac{\alpha_Y}{\alpha_2 + \alpha_Y} = \frac{3/5}{2+3/5}
= \frac{3}{13} = 0.231
\end{equation}
There is a small correction for the running between $M_Z$
and the TeV scale but this is largely compensated by
the two-loop correction and
the agreement remains
as good as for SUSY-GUTS.
This is strong encouragement for the conformality approach.

\bigskip

In the second, non-abelian, example we use $\Gamma = Z_3 \times D_4$ and
choose N=2 to arrive at a unification based on
the Pati-Salam group $SU(4)_C \times SU(2)_L \times SU(2)_R$
instead of the trinification $SU(3)^3$. This is possible
because this non-abelian $\Gamma$ has two-dimensional
representations as well as one-dimensional ones.

The dihedral group $D_4$ consists of eight rotations which leave
a square invariant: two of the
rotations are flips about two
lines which bisect the square and
the other four are
rotations through
$\pi/2, \pi, 3\pi/2$ and $2\pi$ about the perpendicular to the plane
of the square.

In this case the low energy gauge group is thus
$SU(4)^3 \times SU(2)^{12}$. We embed $SU(3)_{color}$ in r
of the $SU(4)$ groups where r = 1 or 2 because r = 3 leads to loss of chirality.
At the same time the $SU(2)_L$ and $SU(2)_R$ are
respectively embedded in
diagonal subgroups of p and q of the twelve $SU(2)$ factors
where p + q = 12.

Since p and q are necessarily integers it is not at all obvious
{\it a priori} that
the value of $\sin^2\theta$ can be consistent with experiment.

The values of the respective couplings at the conformality/unification
scale are now:

\begin{equation}
\alpha_{2L}^{-1}(M_U) = p \alpha_U^{-1}
\end{equation}

\begin{equation}
\alpha_{2R}^{-1}(M_U) = q \alpha_U^{-1}
\end{equation}

\begin{equation}
\alpha_{4C}^{-1}(M_U) = 2r \alpha_U^{-1}
\end{equation}
The hypercharge coupling is related by
\begin{equation}
\alpha_1^{-1} = \frac{2}{5} \alpha_{4C}^{-1} +
\frac{3}{5} \alpha_{2R}^{-1}
\end{equation}

Defining $y = \ln (M_U/M_Z)$ we then find the general expression
for $\sin^2 \theta_W(M_Z)$ to be:
\begin{equation}
\sin^2 \theta_W(M_Z) = \frac{p - (19/12\pi) y \alpha_U}
{p + q + \frac{4}{3} r + (11/6\pi) y \alpha_U}
\end{equation}
Here
\begin{equation}
\alpha_S^{-1}(M_Z) = 2r \alpha_U^{-1} - \frac{7}{2 \pi} y
\end{equation}

Using these formulas and $\alpha_S(M_Z) \sim 0.12$ we
find for the natural choices (for model building) p = 4 and r = 2
that
\begin{equation}
\sin^2\theta_W(M_Z)  \simeq 0.23
\end{equation}
again in excellent agreement with experiment.

\bigskip

It is highly non-trivial that again the gauge coupling unification
works in this case which, according to the lengthy analysis in
the second paper of \cite{FK2}, is the unique accommodation
of the standard model with three chiral families
for all non-abelian $\Gamma$ with order $g \leq 31$.

\bigskip

The successful derivation of $\sin^2\theta_W(M_Z) \simeq 0.23$
from both the abelian orbifold (based on 333-trinification)
and the non-abelian orbifold (based on 422-Pati-Salam unification)
is strong support for further investigation of the detailed phenomenology
arising from the approach.

\bigskip

\section{TeV Unification} 

\bigskip

As one example of this approach arrived at shortly after, but
inspired by, the Oxford conference, let me mention an example
of strong-electroweak unification
at a relatively low ($\sim 4$ TeV) scale\cite{TeV}. 

It was motivated partly by bottom-up
considerations which could be matched to the above top-down idea.

\bigskip

In the standard model, the three couplings are well measured at
the Z-pole, particularly at LEP. The electroweak mixing angle
$\sin^2 \theta (M_Z) = 0.231 = \alpha_Y (\alpha_2 + \alpha_Y)^{-1}$
is close to 1/4 and as the energy scale is raised it increases going through 1/4
at a scale of about 4 TeV. This scale played a role in the 3-3-1 model
\cite{331} and in the more recent study by
Dimopoulos and Kaplan\cite{DK}.

The strong coupling $\alpha_3(\mu)$ relative to the $SU(2)_L$
coupling has a ratio $r(\mu) = \alpha_3(\mu)/\alpha_2(\mu)$
which is $r > 3$ at $\mu = M_Z$ and goes through r=3 at $\mu\sim 400$ GeV
then r=2 at  $\mu \sim 140$ TeV. The value  r = 5/2 is attained at a scale
impressively close to the $\sim4$ TeV scale where $\sin^2 \theta = 1/4$.

We therefore adopt a gauge group $SU(3)^{12}$ at $\mu = 4$ TeV and
identify the trinifiaction gauge groups
$SU(3)_C$, $SU(3)_W$ and $SU(3)_H$ with
2, 5, and 5 of the SU(3) factors respectively.

Assuming all the gauge couplings are equal at this unification scale
there are two predictions: the correct valus of $\sin^2 \theta$
and of $\alpha_3$ at the Z-pole. This is interesting
because usual GUTs predict only one of these two
quantities.

\bigskip

One could stop with such a bottom-up unfication but
the theory becomes more interesting when we marry it
to string orbifolding, this time using
$AdS_5 \times S^5/Z_{12}$. We take N=3 3-branes to achieve
$SU(3)^{12}$. One must then specify the embedding
of $Z_{12}$ in SU(4) such that the scalars are adequate
to allow spontaneous breaking to the standard gauge group.
This leads to the choice ${\bf 4} = (\alpha, \alpha^2, \alpha^3, \alpha^6)$
where $\alpha$ is the 12th root of unity.

\bigskip

The chiral fermions can now be deduced by drawing the dodecagonal quiver
(the nodes are arrange exactly like the numbers on a clock face)
and one finds that there are three chiral families. Actually there are 
five families and two antifamilies, and although there is insufficient
space here to go into technical detail it is possible to relate 
the reason for three families to the difference
between the numerator and denominator in 
the minimalized ratio $r =5/2$, mentioned earlier.
As a final merit of the model it has no GUT hierarchy because there
is no scale above 4 TeV. It is a non-gravitational theory where the Planck scale
is infinite.

It is hoped to pursue the phenomenology of such an approach further
in the future, as well as to investigate the robustness
of the predictions with respect to the input unification scale.

\section{Acknowledgements}

The organizers Steve Abel and Alon Faraggi must be thanked for creating such an excellent
conference and welcoming ambience in Oxford. This work was supported in part by
the US Department of Energy under Grant No. DE-FG02-97ER-41036.

\end{document}